# Casting voids in nickel superalloy and the mechanical behaviour under room temperature tensile deformation

Zhuocheng Xu[a,b], Ben Britton[b], Yi Guo[b*]

a. Department of Aeronautics, Imperial College London, London SW7 2AZ, United Kingdom

b. Department of Materials, Imperial College London, London SW7 2AZ, United Kingdom

**Abstract**

The microstructure of a second-generation nickel base superalloy is studied using X-ray computed tomography (XCT) and scanning electron microscopy (SEM). The as-cast material contains 0.15 (±0.001) vol% voids and these are distributed in the inter-dendritic region. The volume fraction of the voids increases to 0.21 (±0.001) vol% after tensile deformation. Surface observations show evidence of dislocation emissions from the void surface, a mechanism possibly facilitates the expansion of the voids and contributes to the increased void volume fraction. Phenomenological parameters such as stress triaxiality, often believed to control void growth, are investigated through crystal plasticity simulation and compared with literature reported data. The results indicate weak correlation between stress triaxiality and void growth, but this may be possibly due to the lack of data at higher level of plastic deformation, which is limited by the ductility of the material. The distribution of the stress triaxiality field within the sample is heterogeneous and the peak of the triaxiality field is a function of the ratio between notch diameter and sample width. A smaller notch diameter to sample width ratio tend to distribute the triaxiality peaks towards the centre of the sample but also lead to higher strain localisation, an effect that results in early sample failure.

*corresponding author: yi.guo@imperial.ac.uk, kikuchipattern@gmail.com

# 1. Introduction

Superalloys are a class of alloys that are often used to make engineering components work successfully under load at high temperatures. As a result of a better high-temperature behaviour, improved processing routes, and ultimately their affordability, Ni-based superalloys are the most common superalloys that used to make turbine blade for aero-engine [1,2]. During the investment casting of single crystal Ni-based superalloys, many pores can be introduced within the materials during the solidification process and subsequent heat treatment [3,4]. Fundamental understanding of the evolution of pores subject to plastic deformation is therefore of interest.

The formation of the pores includes:

1. the coalescence of secondary dendritic arms leads to insufficient liquid feed perpendicular to the primary dendrite trunk. They are typically found in the intersections between the primary dendrite and the dendrite arms and may be incorporated inside the dendrites [5];
2. the solidification of the residual liquid in-between dendrites facilitated by thermal contraction [5];
3. the partial chemical reduction of the ceramic mould ($SiO_2$ or $Al_2O_3$) or cooling rod by carbon content in the alloy melt, the formation of CO bubbles are mixed in the liquid [4,6];
4. homogenization pores formed during the post casting heat treatment due to asymmetrical diffusion between different regions [3,6].

The first two types of pores are referred to as shrinkage pores.

In the previous studies, by using classical metallography and tomographic analysis, the typical sizes and shapes of these voids are well measured and analysed. In general, shrinkage pores often show irregular shapes with the equivalent diameters of more than 15 µm. Homogenization pores and gas pores often possess spherical shapes and their sizes are about 7-15 µm [3].

The presence of pores may be detrimental to the mechanical properties of the materials as they lead to stress concentration and crack initiation. According to the previous studies, the features such as size, shape, orientation and separation between the pores can all affect the degrees of stress concentration near the pores [6]. Hence, it is important to characterise the distributions, morphology and volume fraction of voids for structural integrity assessment. In earlier work, these features related to voids were often investigated by the combination of classical metallograph and different microscopies. However, as the sizes and the volume fraction of voids are relatively small, the sample preparations for these methods were complicated and the data processing was tedious. As a non-destructive characterisation technique that is able to characterise the 3D interior of materials, X-ray computed tomography (XCT) has become a more effective method to study the pores within the materials [7–11].

The effects of pores to the fracture mechanism of single crystal nickel superalloy varies among different temperature ranges and modes of loading. For uniaxial tensile testing, Xiong et al [12] report that the single-crystal nickel superalloy samples fracture in a quasi-cleavage manner until the testing temperature reaches 650°C. After 850 °C, the rupture surfaces of the sample start to show the dimples, characteristic of ductile fracture. The switch of fracture mechanisms is typically attributed to the increase in the rate of dynamic recovery at high temperatures [12]. As for room temperature fatigue testing, the roles of voids in crack initiations have been reported for both low-cycle and high-cycle fatigue tests [13,14]. In both cases, a single void could serve as a crack nucleation site.

In the present paper, the microstructure of a second-generation nickel superalloy was characterised using XCT and scanning electron microscopy (SEM) before and after tensile testing at room temperature. A crystal plasticity finite element simulation was used to study the heterogeneous deformation field together with the parameters that were often regarded to influence void growth.

## 2. Experimental methods:

The single crystal Ni-based superalloys used in this project was designated DD6, it is a low-cost 2nd generation single crystal alloy. The cost reduction in this alloy was achieved by the replacement of the Re content with refractory substitutes [15]. The compositions of this alloy are shown in Table 1.

*Table 1: The chemical compositions of DD6 alloy*

| Elements | Cr | Co | Mo | W | Ta | Re | Hf | Al | Ni |
|---|---|---|---|---|---|---|---|---|---|
| Wt% | 4.3 | 9.0 | 2.0 | 8.0 | 7.5 | 2.0 | 0.1 | 5.7 | 61.4 |

The tensile specimens were machined by electron discharge machining (EDM) with the loading direction parallel to the [001] crystal direction and the direction normal to the sample surface nearly parallel to the [110] direction. The dimension of the testing specimen is shown in Fig. 1. The sample thickness is 1 mm and the middle of the sample was notched with a 5 mm diameter notch, leaving 1 mm width at the smallest sample cross section. The reason for the notch is to confine deformation within the central region to ensure a consistency of XCT scans to capture the same microstructure before and after deformation. Besides, the four circular holes near the gripping region are drilled so that pins can be applied to align the specimen with the tensile axis. The samples were mechanically polished using 400, 800, 1200, 2000, and 4000 grit papers and finished with 1:1 water diluted colloidal silica polishing to a scratch free surface finish.

The sample was mechanically tested in tension under displacement control with an extension rate of 0.1 mm/min up to a 0.38 mm displacement where the sample is close to fracture based on a separate test.

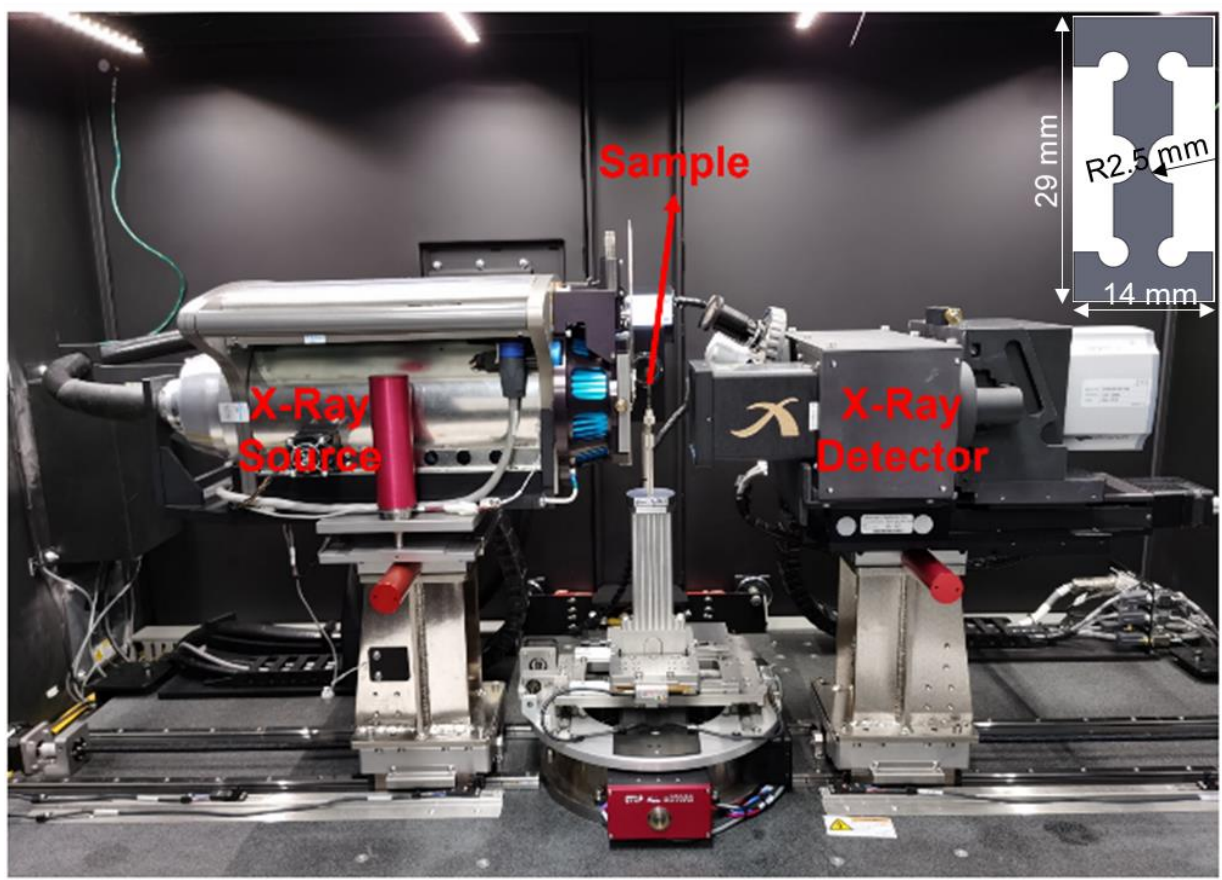

*Figure 1: XCT scan set up and the sample geometry*

The statistics of the shape and size distribution of the voids before and after deformation were captured by XCT scans. The tomographic scans were performed using Zeiss Versa XRM-520 (Fig. 1) operating at 160 kV and 62 µA with a sample to source and sample to detector distance of 13 mm and 30 mm respectively. Radiographic projections were collected using a x4 objective lens and 10s exposure time. 2520 projections were collected, and the 3D object was reconstructed from these projections using a filtered back projection algorithm. The duration for the scan was 7 hours and the given scanning parameter resulted in a 0.68 µm$^3$ voxel size. The region of interest was chosen to be as close to the centre of the notched region as possible and the size of the scanned region was 2400 µm x 2400 µm x 1950 µm. Density variations in the material give rise to the differential absorption of X-ray and allow for the segmentation of features in the sample. This task was performed using Avizo.

## 3. Results

The features giving rise to the density variation in the material are revealed from the as-collected radiographic projections, an example is given in Fig. 2. Due to the differences in density and therefore the rate of X-ray absorption, three features stand out from the radiographic image, namely voids, dendrites, and inclusions. Energy dispersive spectroscopy (EDS) indicates the inclusions are Si rich particles (shown in the Supplementary figures).

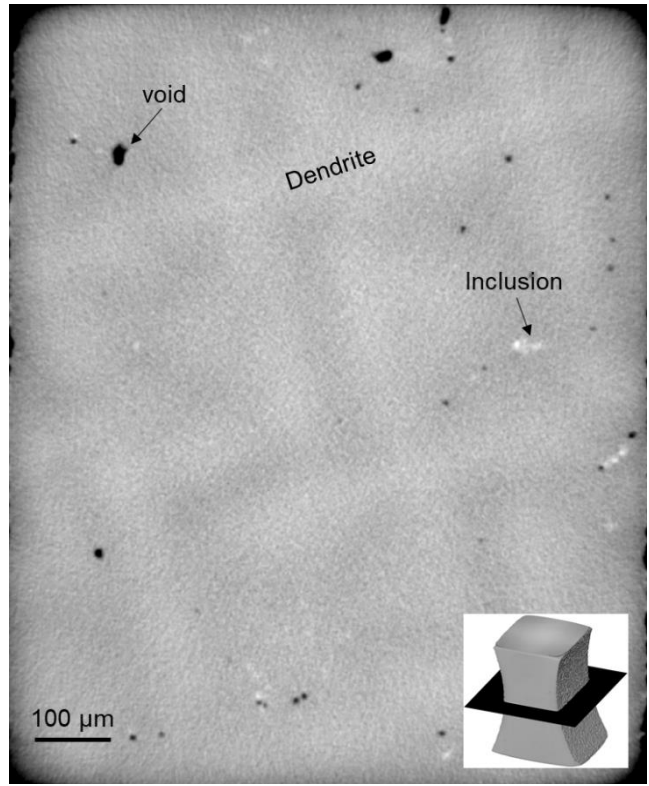

*Figure 2: A cross sectional radiography image (location indicated by the inset) showing the microstructures of the sample include: voids, inclusions, and dendrite.*

A 3D view of the microstructures can be generated by image segmentation based on the grey scale contrast and the results are shown in Fig. 3. The opacity histogram of the matrix (appears in grey) was adjusted to bring out the contrast of features inside. The dendrites are shown in green and the voids in red. Due to close grey scale contrast between inclusions and the dendrites, as can be seen from Fig. 2, segmentation of the inclusions involves significant amount of laborious manual intervention, therefore they are shown in the small volume of interest analysis in Fig. 4, and not the whole volume that was rendered as in Fig. 3. It can be seen from the top-down view of Fig. 3b that the voids are distributed in between the dendritic arms, characteristic of shrinkage pore formation and consistent with literature reports [16–18].

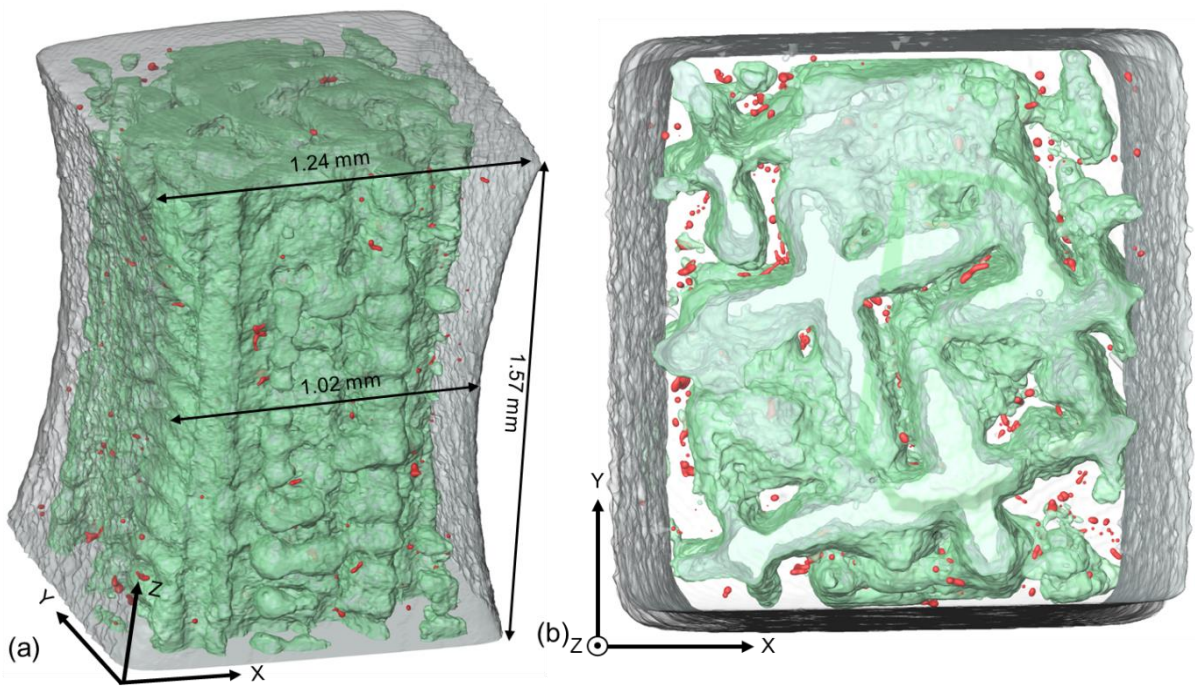

*Figure 3: A 3D visualisation of the scanned region before deformation with (a) 3D view and (b) top-down view. The grey contrast represents the matrix while the green and red colours represent the dendrites and the voids respectively.*

In order to quantify the initial porosity and its evolution after deformation, a rectangular volume was extracted from the centre of the scanned region. This was done by selecting two reference voids along each of the X-Y-Z axis. The six reference voids were found in the undeformed sample and correlated in the deformed sample. This method enables an accurate assessment of the void statistics within the same volume and the results are shown in Fig. 4. Both the equivalent diameter and sphericity were analysed using the label analysis function in Avizo, with the respective formulation of Equivalent Diameter = $\sqrt[3]{6V/\pi}$, Sphericity = $A^3/36\pi V^3$, where A and V are the surface area and the volume of the object respectively. The porosity was calculated by dividing the sum of the void volumes inside the box by the volume of the box. The results show that the initial porosity of the sample is ~0.15% which increased to ~0.21% after deformation (to a strain just before fracture). Most voids have equivalent diameters in the 5 μm to 10 μm range and the biggest void is around 40 μm. The shapes of the voids do not seem to change significantly after deformation as can be seen from the sphericity histogram from Fig. 4. The inclusions are coloured red in Fig. 4, and they consist of ~0.04% volume fraction of the material. From the top-down view of Fig. 4 it seems that the inclusions tend to cluster with the voids and therefore they are also distributed in-between the dendritic arms.

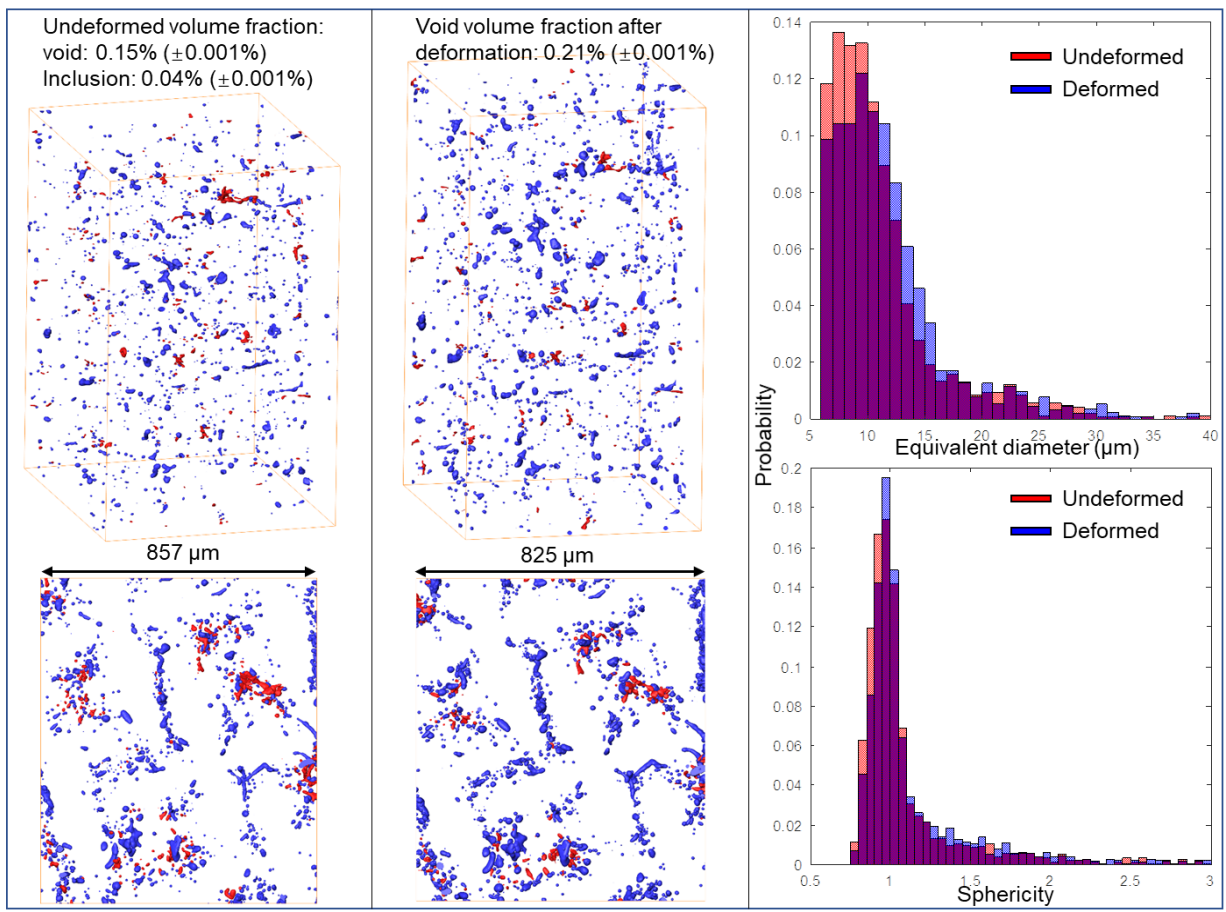

*Figure 4: The distribution of voids and inclusions in the volume of interest before and after deformation. The histograms indicate the size and shape distribution of the voids.*

To investigate the void size change in detail, five voids were selected from different locations of the sample, which each having different local deformation conditions due to the sample geometry. The locations of the voids are indicated in Fig. 5 together with a table showing the size evolution after deformation. Voids 1 and 4 are located at the top and bottom of the shown volume towards the centre of the cross sections. Voids 2 and 3 are located near the central cross section with the former at the centre of the volume and the latter towards the notch surface on the right-hand side of the sample. Void 5 is located on the front surface and the deformation around this void, corroborating its size evolution, is shown by the scanning electron microscopy (SEM) image in Fig. 5 c and d. The growth of the void is probably facilitated by the emission of two slip systems from the top of the void as indicated by the SEM image (Fig. 5d). The growth ratios among the 5 voids don't seem to show a particular trend with respect to their position in the sample.

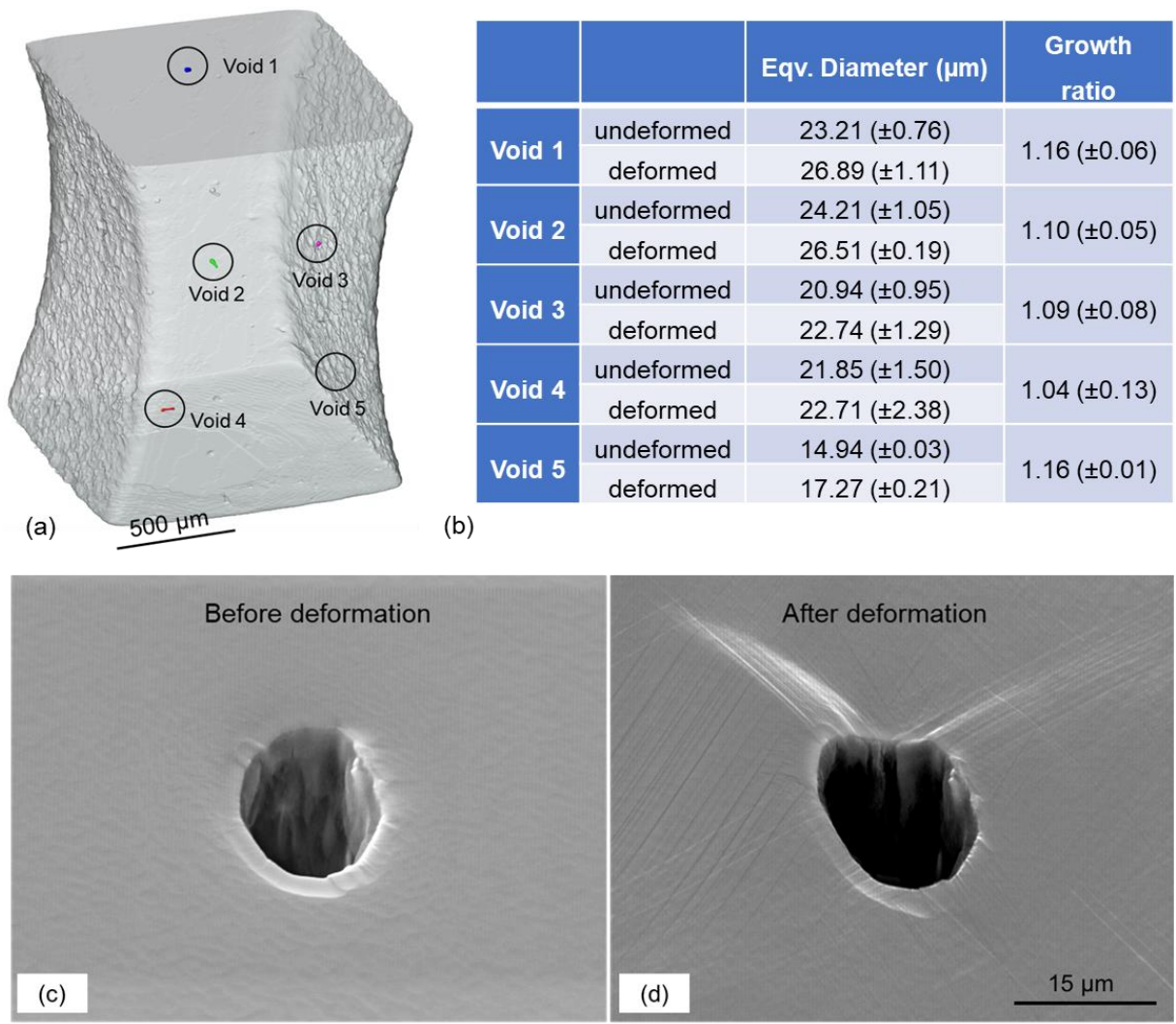

Figure 5: The location of a few selected voids for further analysis (a), their size before and after deformation (b), and the SEM images of a surface void (void 5) showing dislocation activity facilitating void growth (c) and (d).

The role of the voids in the room-temperature fracture process may be inferred by looking at the fracture surface. A scanning electron microscopy derived fractography is shown in Fig. 6. It can be seen that the sample fails by cleavage fracture evidenced by the presence of the nearly smooth facets. This is in contrast to a ductile fracture where a dimpled fracture surface is seen due to void growth and coalescence. The voids are seen on the cleavage surface, indicated by the circles, and they are not associated with visible local microcracks. No strong evidence links the voids to either the nucleation or the propagation of cracks.

Instead, the crack seems to have nucleated from the two tetragonal protrusions indicated by the dashed boxes. On the one hand, these tetragonal protrusions are located in the middle of the sample gauge and close to the edge of the sample, where a digital image correlation study (shown in Supplementary figures) indicate significant strain localisation. On the other hand, the close resemblance to Thompson tetrahedron may suggests dislocation reactions on adjacent (111) planes to form a sessile $<0\bar{1}\bar{1}>$ type dislocation on (100) plane [19], i.e. a Lomer Cottrell lock. The strength of the lock has been predicted to be on the order of 10 GPa through molecular dynamic simulations [20]. Therefore, the local plastic deformation in the middle of the gauge section can no longer be supported by the emission of dislocations. This may have led to the initial crack nucleation. This hypothesis needs further experimental confirmation and will not be analysed further in this paper.

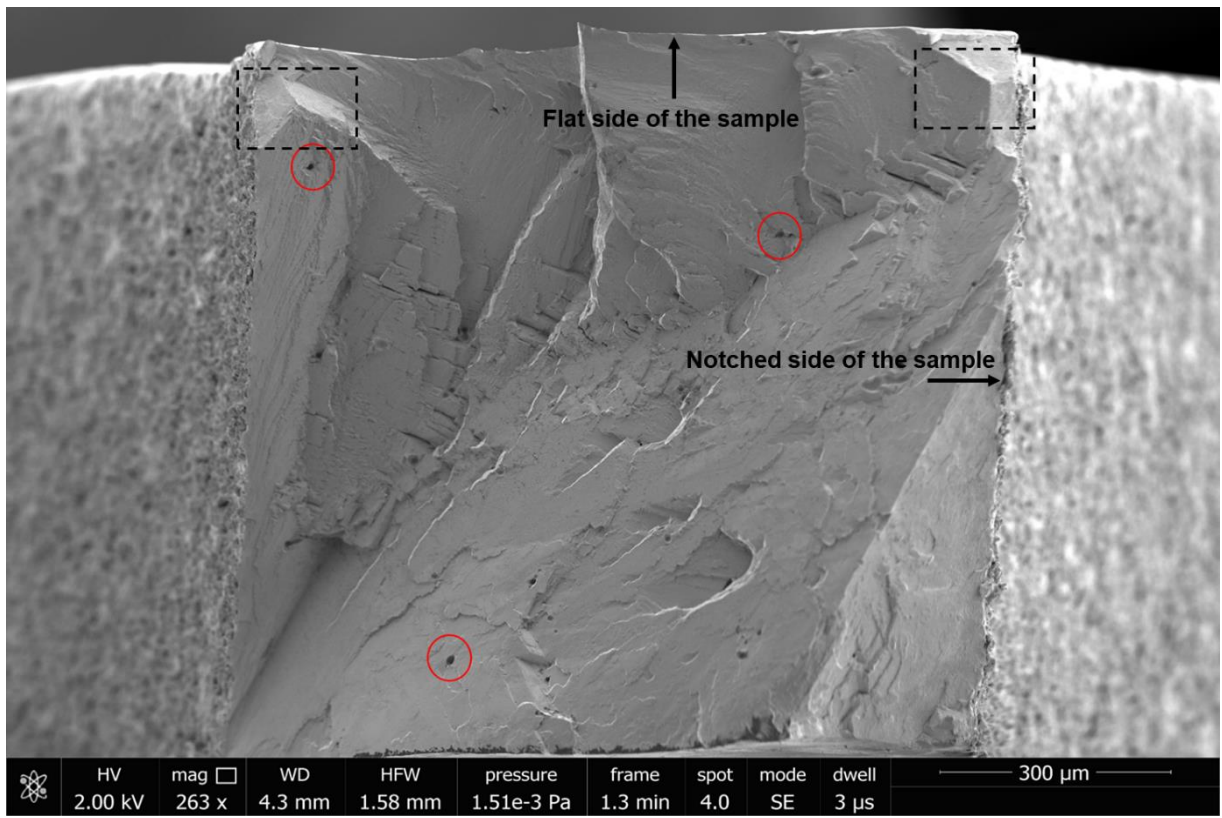

*Figure 6: The secondary electron mode SEM micrograph for the fracture surface after tensile testing. The sides cut by EDM are the left- and right-hand sides of the image. Within the figure, the likely crack initiation regions are labelled with black dashed squares. The voids within red circles show no significant influences on crack growth. In this image, the loading direction is out-of-page.*

## 4. Discussion

The void volume fraction in the as cast alloy has been studied by XCT. The porosity of the as cast material is ~0.15% and increased to ~0.21% during tensile testing before fracture. Fundamentally, the growth of voids may be supported by dislocation activities around the void, which may involve one or more of the mechanisms involving the emission of prismatic loops [21–24] and the glide of anti-parallel dislocation pairs [25–27]. The small increase of equivalent diameter shown in Fig. 7 is likely due to the glide of a dislocation pair as evidenced by the two sets of slip bands emitted from the top of the void. The void growth in this alloy, subjected to room temperature deformation, is different to a more homogeneous dislocation slip and higher void growth rates that have been seen in other materials such as Cu [28] and Al [29] during tensile testing.

On a higher order length scale, the growth of voids has been phenomenologically linked to stress triaxiality (the ratio between hydrostatic stress and equivalent stress) and strain [30,31]. For the five voids indicated in Fig. 5, the local strain and stress triaxiality was studied using a physically based crystal plasticity simulation. The methods and the underlying theories for this model were introduced by Dunne et al [32], and a brief description is given in the following:

The total deformation gradient $\boldsymbol{F}$ is decomposed into an elastic ($\boldsymbol{F}^e$) and a plastic deformation gradient tensor ($\boldsymbol{F}^e$) as

$$\boldsymbol{F} = \boldsymbol{F}^e \boldsymbol{F}^p \quad (1)$$

The plastic velocity gradient ($\boldsymbol{L}^p$) is determined from dislocation slip rate ($\dot{\gamma}^\alpha$) on the $\alpha^{th}$ slip system by

$$\boldsymbol{L}^p = \dot{\boldsymbol{F}}^p \boldsymbol{F}^{p-1} = \sum_{\alpha=1}^{12} \dot{\gamma}^\alpha \boldsymbol{s}^\alpha \otimes \boldsymbol{n}^\alpha \quad (2)$$

Where $\boldsymbol{s}^\alpha$ and $\boldsymbol{n}^\alpha$ are the slip shear (Burgers vector) direction and slip plane normal direction of the $\alpha^{th}$ slip system respectively. The slip rate ($\dot{\gamma}^\alpha$) is given by

$$\dot{\gamma}^\alpha = \rho_m b^{\alpha 2} v exp\left(-\frac{\Delta F}{kT}\right) sinh\left(\frac{(\tau^\alpha - \tau_c^\alpha)\Delta V^\alpha}{kT}\right) \quad (3)$$

Where $\rho_m$ is the mobile dislocation density, $b^\alpha$ is the Burgers vector magnitude of the $\alpha^{th}$ slip system, $v$ is the jump frequency of dislocations to overcome obstacles, $k$ is the Boltzman's constant, $T$ the thermal dynamic temperature. $\tau^\alpha$ is the resolved shear stress on the $\alpha^{th}$ slip system and mobilizes dislocations once it exceeds a critical value, $\tau_c^\alpha$. The rate sensitivity is controlled by the activation energy, $\Delta F$, and the activation volume, $\Delta V^\alpha$.

The Taylor hardening model is employed to account for the hardening of slip systems due to the accumulation of dislocations. Therefore, the evolution of the critical resolved shear stress ($\tau_c^\alpha$) is given by

$$\tau_c^\alpha = \tau_0^\alpha + Gb\sqrt{\rho_{SSD} + \rho_{GND}} \quad (4)$$

Where $\tau_0^\alpha$ is the initial lattice friction of the slip system $\alpha$, $G$ is the shear modulus, and $\rho_{SSD}$ is the statistically stored dislocation density whose evolution is a linear function of the effective plastic strain rate ($\dot{p}$)

$$\dot{\rho}_{SSD} = \lambda \dot{p} \quad (5)$$

in which $\lambda$ is an isotropic hardening coefficient.

The effective plastic strain rate ($\dot{p}$) is linked to the plastic deformation rate ($\boldsymbol{D}^p$) as

$$\dot{p} = (\tfrac{2}{3}\boldsymbol{D}^p : \boldsymbol{D}^p)^{1/2} \quad (6)$$

Where $\boldsymbol{D}^p$ is obtained by

$$\boldsymbol{D}^p = \text{sym}(\boldsymbol{L}^p) \quad (7)$$

The GND density is determined by relating the Nye dislocation tensor ($\Lambda$) with lattice curvature

$$\boldsymbol{\Lambda} = \text{curl}(\boldsymbol{F}^p) = \sum_{\alpha=1}^{12} \rho_s^\alpha \boldsymbol{b}^\alpha \otimes \boldsymbol{s}^\alpha + \rho_{en}^\alpha \boldsymbol{b}^\alpha \otimes \boldsymbol{n}^\alpha + \rho_{em}^\alpha \boldsymbol{b}^\alpha \otimes \boldsymbol{m}^\alpha \quad (8)$$

Where the $\rho_s^\alpha$, $\rho_{en}^\alpha$, and $\rho_{em}^\alpha$ represent, respectively, the screw dislocation along the shear direction ($\boldsymbol{s}^\alpha$), edge dislocation along slip plane normal ($\boldsymbol{n}^\alpha$), and edge dislocation along the $\boldsymbol{m}^\alpha$ (= $\boldsymbol{s} \times \boldsymbol{n}$) direction of the $\alpha^{th}$ slip system. The plastic deformation was homogenized into the slip of 12 FCC slip systems and equation 8 is solved using a L2-norm minimization. GND density is obtained by

$$\rho_{GND} = \sqrt{\sum_{\alpha=1}^{12}((\rho_s^\alpha)^2 + (\rho_{en}^\alpha)^2 + (\rho_{em}^\alpha)^2)} \quad (9)$$

The above formulations are implemented through a user-material (UMAT) subroutine using ABAQUS standard analysis. The finite element model was meshed with 5 μm element size at the central notched region and increased to 100 μm towards the shoulder of the sample, using three-dimensional, 20-noded quadratic hexahedral elements with reduced integration (C3D20R). The model (consists of 68910 elements with 297586 nodes) together with the loading and boundary conditions is shown in Appendix Figure 3. Frost and Ashby has proposed an estimation of the activation energy by $\Delta H = \omega G b^3$, where $\omega$ is a constant depending on the obstacle strength, G is shear modulus, and b is burger's vector. Taking b = 2.54 x 10-4 µm, G = 90GPa, and $\omega$ =0.024 for rate insensitive deformation at room temperature [33], $\Delta H$ is then 3.456 x 10-20 J/atom. The activation volume has the form of $\lambda_p b d$ where d is the length scale of the activation event, which is approximately the burger's vector length. Taking the obstacle spacing, $\lambda_p$, as 11.6b [34] the activation volume, ΔV, is then 11.6 b3. The dislocation jump frequency, $\nu$, is taken to be 1 x 1011 s-1 [35], smaller than the Debye frequency of 1 x 1013 s-1 [36] in order to minimize rate sensitivity. A detailed description of the simulation parameters, as summarized in Table 1, can be found in [32,33,37]. The constitutive equations have been applied to a variety science and engineering relevant investigation including fatigue crack nucleation [38–40], deformation anisotropy in polycrystal deformation [41–43], galling wear [44,45], and so on.

Table 1: Single crystal elastic constants and simulation parameters.

| ρ (m-2) | b (µm) | $\nu$ (s-1) | ΔH (J/atom) | T (K) | $\tau_c$ (MPa) | λ(m-2) | ΔV |
|---|---|---|---|---|---|---|---|
| 1x10^10 | 2.54 x 10-4 | 1.0 x 10^11 | 3.456 x 10-20 | 293 | 450 | 3.0 x 10^14 | 11.6b3 |

Elastic constants: $C_{11}$=252 GPa, $C_{12}$=161 GPa, $C_{44}$=131 GPa

The simulated load-displacement curve was brought into agreement with the experimental curve by adjusting the hardening parameter ($\lambda$) and the critical resolved shear stress ($\tau_c$). The computed stress-strain response obtained from the model is shown in Appendix Figure 4, demonstrating reasonable agreement with experimental data.

The simulation results, i.e. the stress triaxiality and strain, are shown in Fig. 7 and 8 respectively.

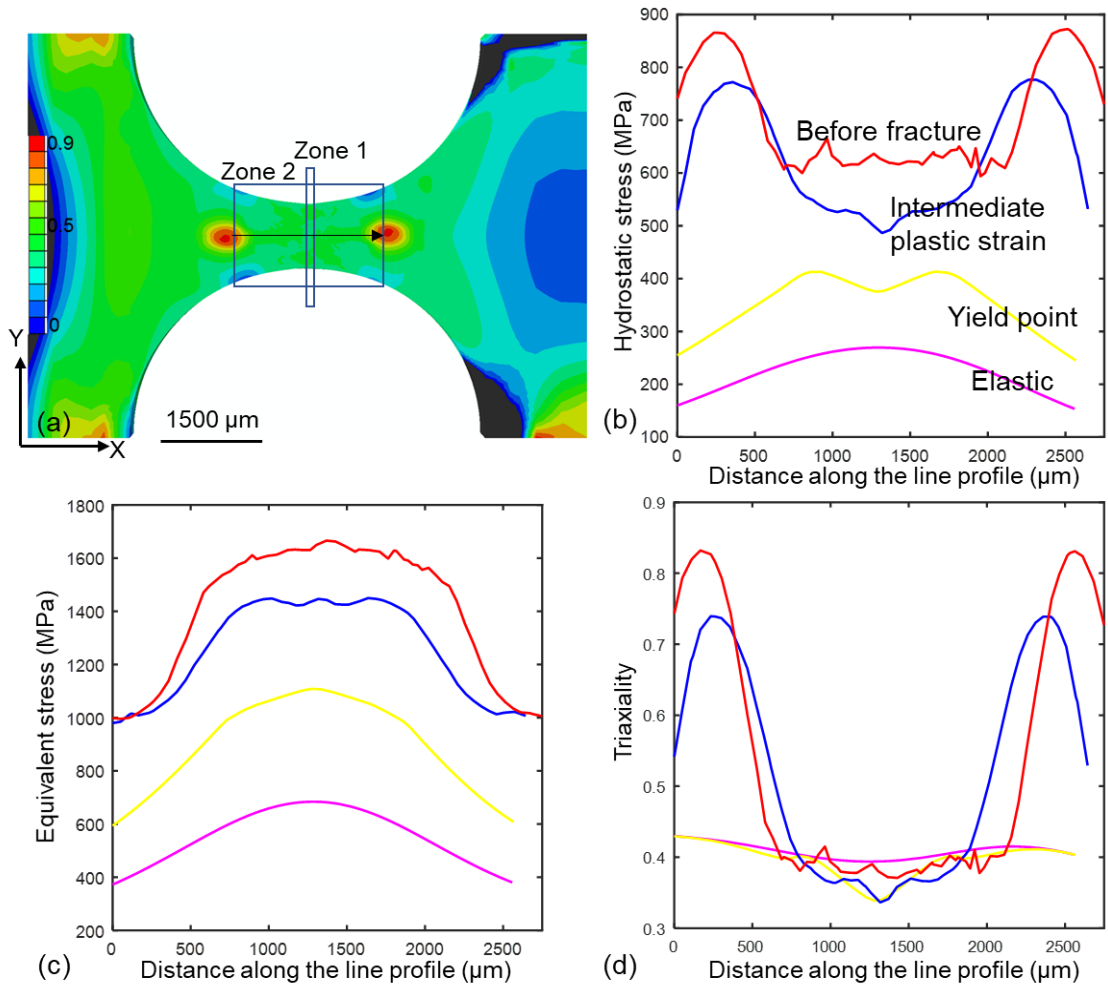

*Figure 7: (a) Stress triaxiality field from the middle XY plane of the sample; (b), (c), and (d) show, respectively, the hydrostatic stress, Von Mises equivalent stress, and stress triaxiality along the arrow direction indicated in (a).*

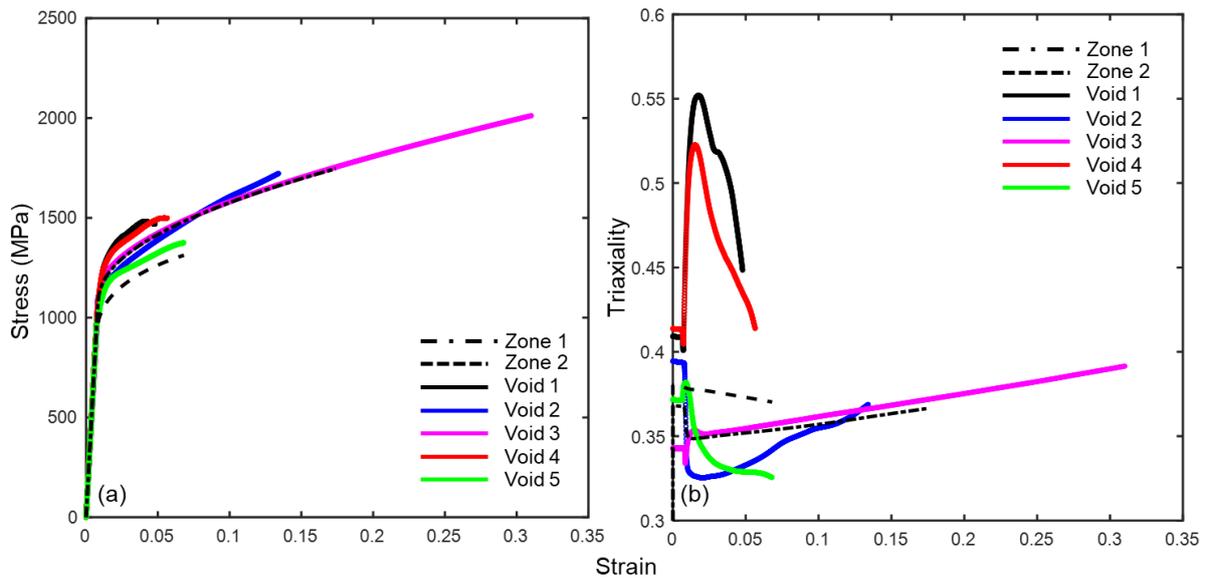

*Figure 8: Crystal plasticity derived simulations of stress-strain curve (a) and stress triaxiality*

*(b) at the location of the five voids indicated in Fig. 5. The dashed curves show the variation of properties of the curved gauge sample, averaged over the zones indicated in Fig. 7.*

It can be seen from Fig. 7b that the hydrostatic stress reaches a maximum value towards the middle of the sample gauge during the elastic loading. After inception of plasticity the magnitude of hydrostatic stress in the centre continue increasing but the peak value tends to shift towards the shoulders of the gauge section. On the other hand, the equivalent stress peaks at the centre of the sample (Fig. 7c) and the distribution persists throughout the deformation. The patterning of the hydrostatic stress and the equivalent stress fields leads to lower triaxiality stress field in the middle of the sample and the opposite towards the shoulder, as visually evident from Fig. 7a. The advantage of the notched sample gauge is that deformation and failure is constrained at the middle, facilitating observations both in the SEM and XCT. However, the variation in shape profile highlights that the majority of deformation within the sample are not uniaxial and therefore it is difficult to present a representative stress-strain curve. This is illustrated by the stress strain curves averaged over Zone I and Zone II (indicated in Fig. 7a) shown by the dashed lines in Fig. 8a. the stress-strain curves are different, where gradually increasing cross sectional area reduces the applied stress thus limiting slip and it leads to lower strain.

This heterogeneity is important when we consider how these boundary conditions impact the local stress strain behaviour at and near the voids, as indicated in Fig. 5a is shown in Fig. 8a, and the corresponding strain-triaxiality plot in Fig. 8b. The variations of the stress-strain curve are due to both heterogeneous slip band formation and the geometric factors of the sample. Void 3, which is located in the middle of the gauge section towards the surface of the sample, shows the highest local strain. Note that the sample fractured by cracking at the edge of the sample in the middle of the gauge. Voids 1 and 4, which are located further away from the middle towards the shoulder, show significantly lower local strain. The drop of stress triaxiality for void 1 and 2 is due to stain localisation at the centre of the sample gauge after the onset of plasticity, leading to a relaxation of the hydrostatic pressure (in this case, a drop of the S22 component) towards the shoulder of the gauge.

To understand the significance of the strain and stress triaxiality in void growth in the current investigation, some literature data [10,46,47] are reproduced and compared in Fig. 9a. For this literature data the stress triaxiality labelled on Fig. 9a were estimated from the shape of the samples based on the Bridgman equation [48,49]. The range represents the stress triaxiality evolution in the beginning and the end of the tensile testing. In the literature data, these experiments were performed using synchrotron radiation where the high X-ray flux enables quicker tomographic scans and therefore higher time resolution [50], whereas the current investigation was performed using laboratory XCT and only a pre-deformation and a post-deformation scans were made. The literature data suggest that void growth in some materials, notably 5741 aluminium alloy, have an 'incubation' period up to 0.3 strain beyond which void growth rate increased significantly, while in some other materials such as martensitic steel and 2024 aluminium alloy, the voids grow from the outset of deformation. At lower strain level, i.e. < 0.3, it seems that at a given strain, pure copper demonstrates higher void growth rate than 5741 aluminium despite the stress triaxiality been lower. Similarly, the martensitic steel and the 2024 aluminium alloy demonstrate similar void growth trend, even under very different triaxiality conditions.

Comparing this data with our experiment results reveals that void 3 in the current investigation shows lower void growth ratio compared to copper at 0.3 strain despite a similar triaxiality condition, and the void 4 shows lower growth ratio at 0.05 strain compared to 2024 aluminium even though the triaxiality is higher. The local strain and stress triaxiality evolutions in Fig. 8 were used to predict void growth based on a Rice and Tracy model modified by Huang [30,31]:

$$\frac{dR}{R} = 0.427 T^{1/4} \exp(\frac{3}{2}T) d\varepsilon \quad (8)$$

where $dR$ is void radius increment, $R$ is the current void radius, $T$ is the stress triaxiality, and $\varepsilon$ is the equivalent plastic strain. The predicted void growth curves are shown in Fig. 9b. It looks that equation 8 fails to predict the void growth in the current material. However, the void growth in the current investigation is around the spatial resolution of the XCT scan, which lead to uncertainty, as reflected by the error bar (estimated from varying the grey scale threshold during void segmentation), on the true gap between the measured data and the predicted curve. To test the efficacy of equation 8 more data points in the higher void growth ratio regime are needed but this will not be available for the current material under room temperature deformation limited by the ductility.

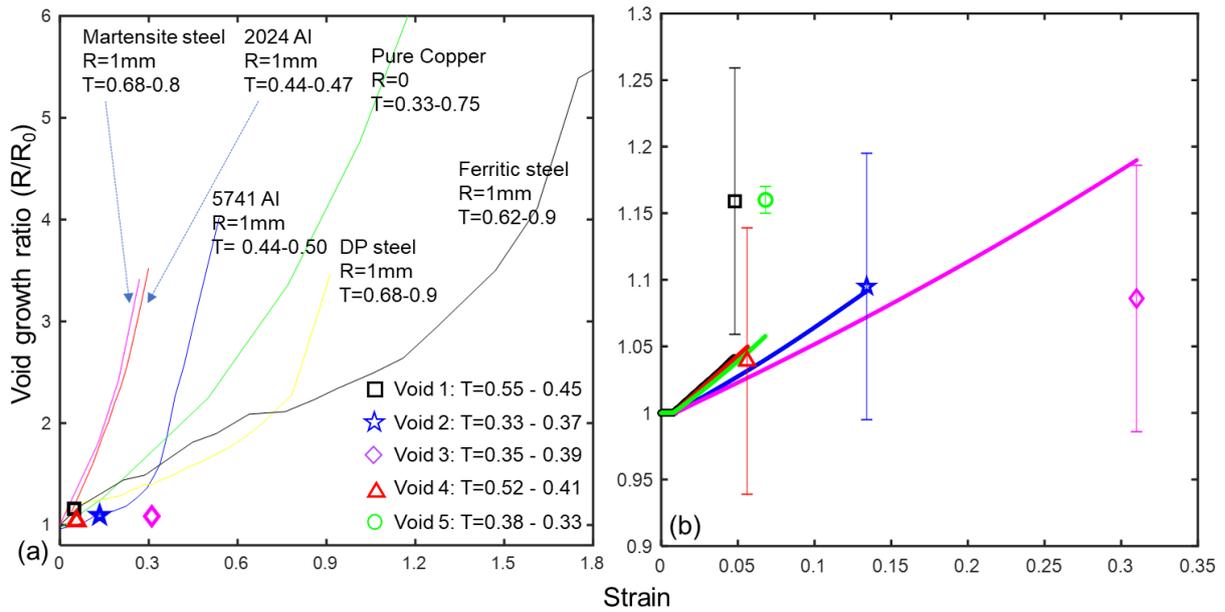

Figure 9: (a) Reproduced literature data regarding void growth ratio (current equivalent diameter/initial equivalent diameter) as a function of strain under the influence of triaxiality loading; (b) predicted void growth curve compared with experimental measurement.

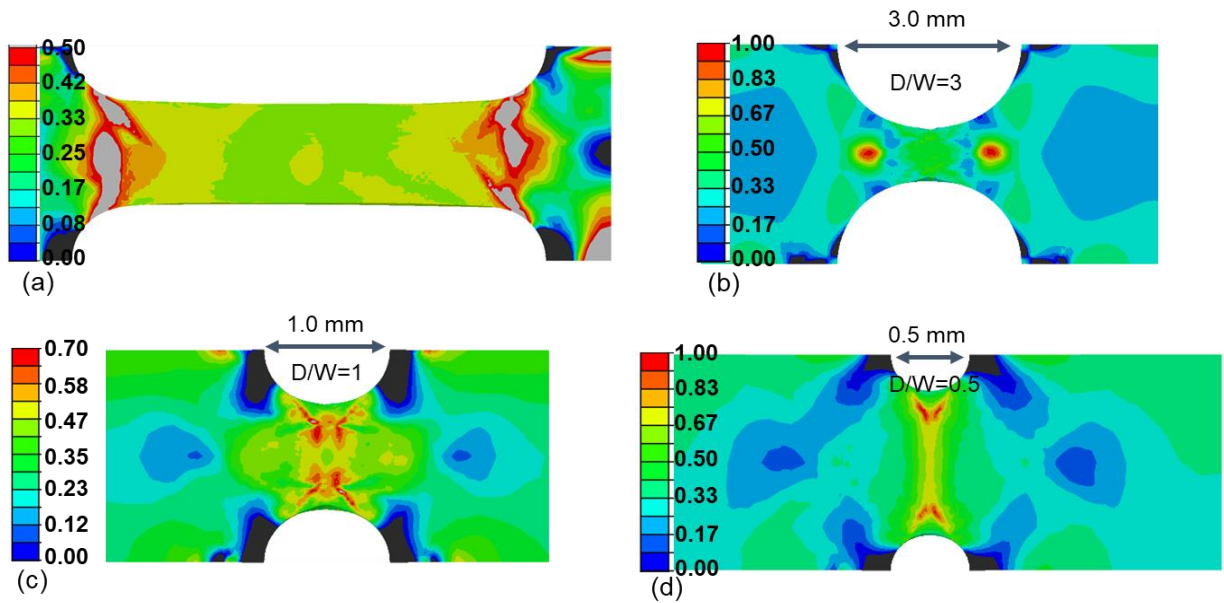

Figure 10: Stress triaxiality distribution in samples with varying notch dimensions.

The Bridgeman equation assumes the triaxiality field is in the centre of the notched sample, but this is not necessarily true and can be modified with notches. The stress triaxiality field in Fig. 7 indicate that for the current sample design with notch radius of 2.5 mm and 1 mm notch separation, the triaxiality peak is not at the sample centre. To explore how the notch size affects the triaxiality distribution, multiple notch-to-separation ratios (D/W, the ratio between notch diameter and notch separation) were simulated and the results shown in Fig. 10. Note that for all the simulations the notch separation (sample width in-between the bottom of notches) are kept constant at 1mm. For the straight sample, the stress triaxiality in the gauge section is relatively uniform and close to 0.33, in agreement with the Bridgeman equation [49], and high triaxiality peaks appear at the shoulder of the sample. When a notch is introduced, significant heterogeneity in the distribution of the triaxiality field can be seen (see Fig. 10). As the notch-to-separation ratios decrease, triaxiality peaks shift towards the centre of the sample and the triaxiality field seem to be confined in the centre of the sample when the notch-to-separation ratio is ≤1. The strain accumulated in the mid-section of the sample gauge (Zone I, Fig. 7a) for the samples with varying D/W at the same remote displacement are compared in Fig. 11.

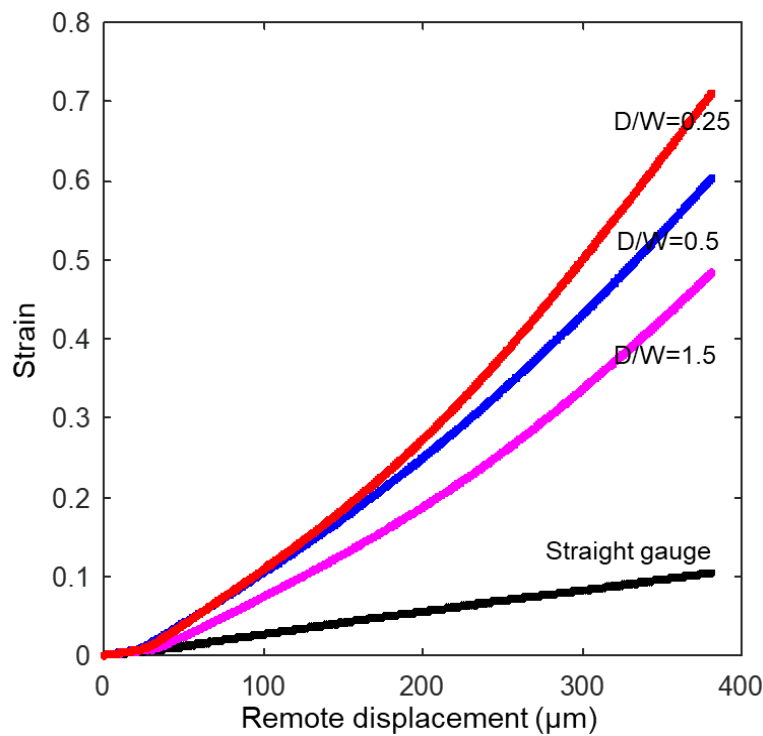

*Figure 11: Comparison of the strain localisation at the middle of the sample gauge for different sample notch design. D/W stand for the ratio between the notch diameter and the smallest sample width across the notched gauge section.*

Samples with smaller notch diameter to sample width ratio develop higher local strain concentration for the same remote displacement. This is to say that for a given ductility of a sample the one with smaller D/W ratio would fracture earlier, yet a common practice to test the effect of increased stress triaxiality to fracture is to notch the sample with smaller notch size. In this scenario, it is difficult to conclude whether it is strain or stress triaxiality that promote early fracture of the sample. This implies that understanding here may not be driven by experiments in the phenomenological parameter space and this motivates a bottom up approach, where some recent molecular and dislocation simulations proposed promising results [25,51].

**Conclusion**

The microstructure of a second-generation Ni base superalloy was characterized by XCT. The results reveal the as cast material contains ~0.15 vol% porosity and ~0.04 vol% inclusions. The porosity increased to ~0.21 vol% after room temperature uniaxial tensile deformation. The voids show small increase in size possibly facilitated by the emission of dislocations from the void surface. Crystal plasticity finite element simulation was used to understand the triaxiality and strain distribution in the sample and their influence on void growth. While the current results suggest the Rice and Tracy model failed to predict the void growth, limited ductility in the sample makes it challenging to understand how this model could be improved. Comparison of literature data regarding void growth seem to suggest that, physically, there are other parameters (e.g. Image force) influencing void growth apart from stress triaxiality and strain. The simulation results suggested that for the high stress triaxiality field to be distributed in the centre of the notched sample gauge, the notch diameter to notch separation ratio should be smaller than 1.

**Author contribution**

This work was conducted in the Imperial College Materials Department while ZX was reading for an MEng degree. ZX performed the XCT scan, analysed the XCT results, and drafted the introduction of the paper. YG analysed the XCT results, performed crystal plasticity simulation and drafted the rest of the paper. TBB and YG supervised the project.

**Acknowledgement**

TBB acknowledges funding of his research fellowship. The authors would like to acknowledge the support from Beijing Institute of Aeronautical Materials (BIAM) where Dr Zong Cui manufactured the material used in this study. The research was performed at the BIAM–Imperial Centre for Materials Characterisation, Processing and Modelling at Imperial College London. The microscope and loading frame used to conduct these experiments were supported through funding from Shell Global Solutions and provided as part of the Harvey Flower EM suite at Imperial. We wish to acknowledge the support of the Henry Royce Institute for ZX through the Royce PhD Equipment Access Scheme enabling access to Zeiss Versa 520 facilities at Royce@Manchester; EPSRC Grant Number EP/R00661X/1. The help from Mr Daniel Sykes in setting up the XCT scans is gratefully acknowledged. We thank Prof Fionn Dunne for helpful discussions about this work.

**Data Availability**

The raw X-ray tomography data can be found at Zenodo under the same title as this publication (DOI: https://doi.org/10.5281/zenodo.4432446).

Supplementary Figures

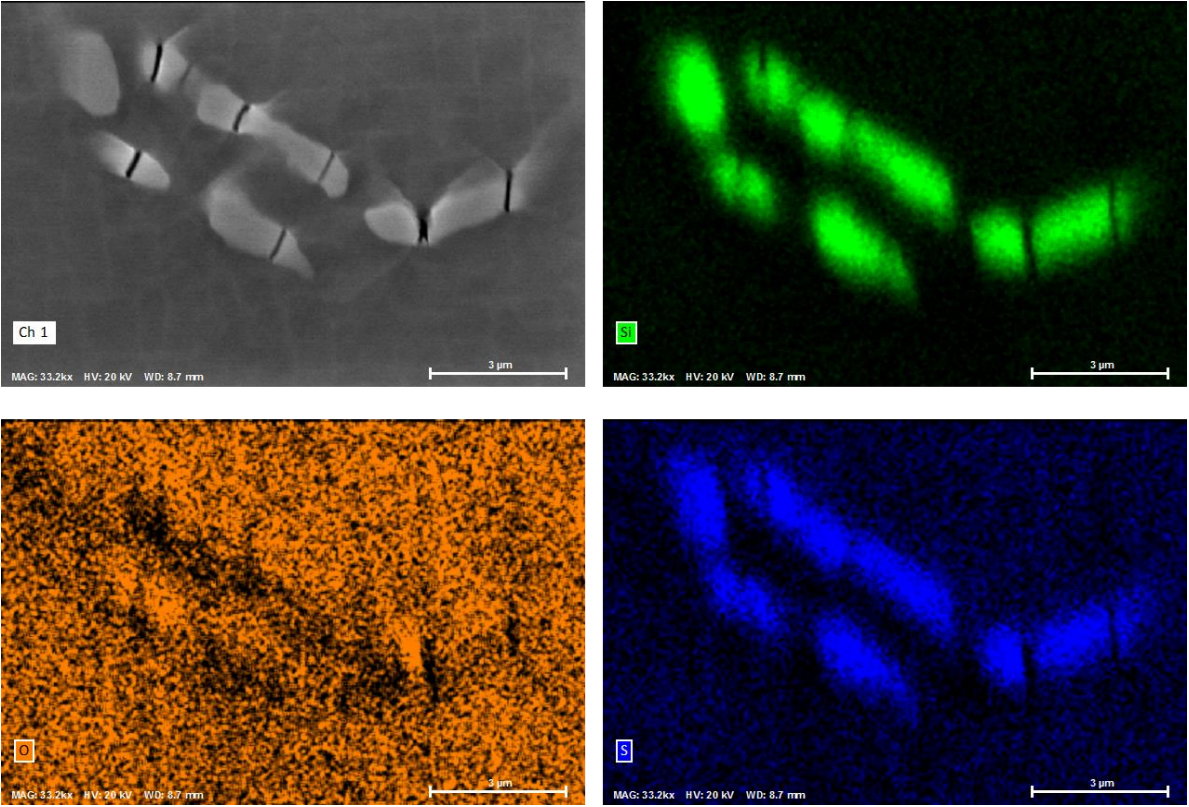

*Supplementary Figure 1: EDS analysis show the inclusions in the material are high in Si and S concentrations.*

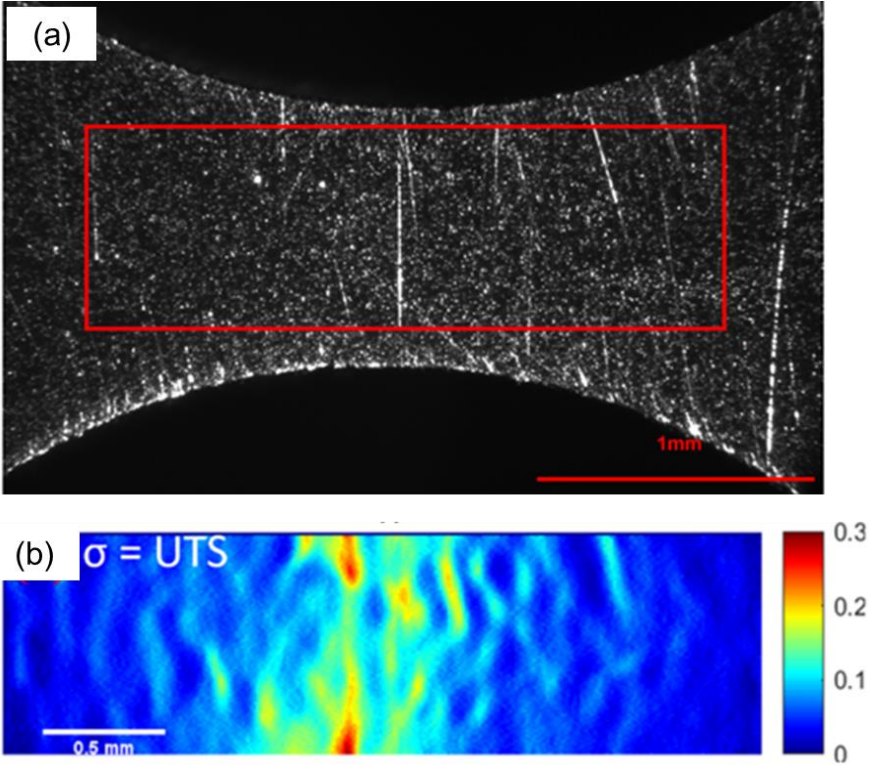

*Supplementary Figure 2: (a) Sample surface patterned with carbon particles for digital image correlation (DIC). The red box indicates the region of interest for analysis. (b) DIC results showing the distribution of $\varepsilon_{11}$. Strain accumulation at the edge of the sample agrees with simulation and correlates with the fracture site.*

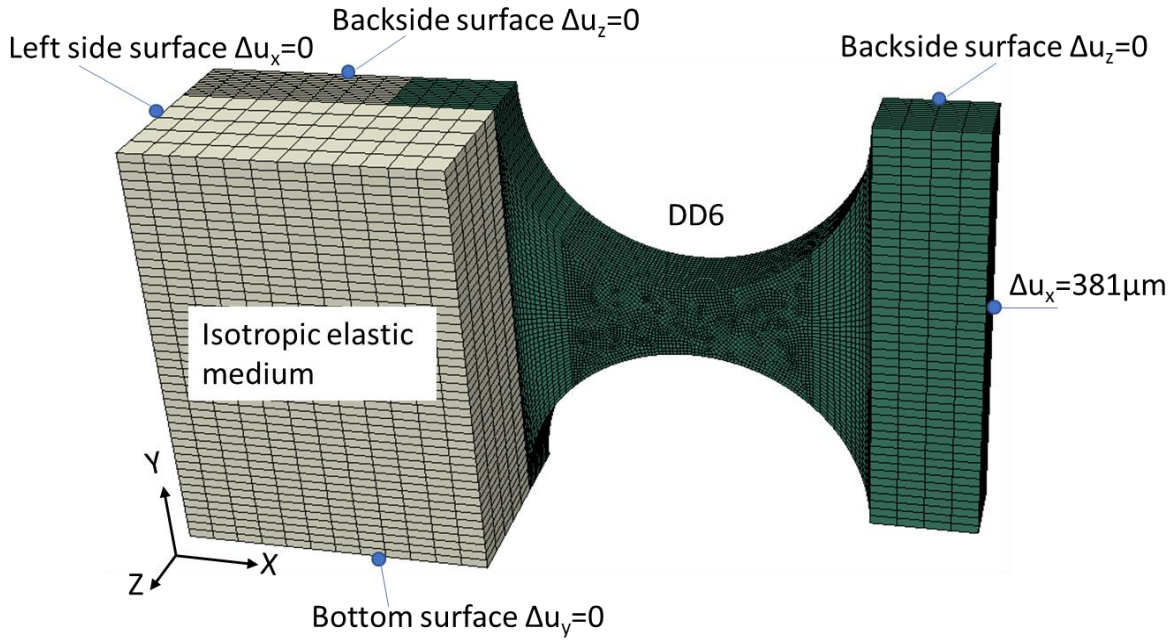

*Supplementary Figure 3: Schematics of the FE model. An isotropic elastic medium is attached to the left-hand side of the sample to simulate the machine compliance. Boundary conditions are indicated where displacement is enforced on the right-hand side surface.*

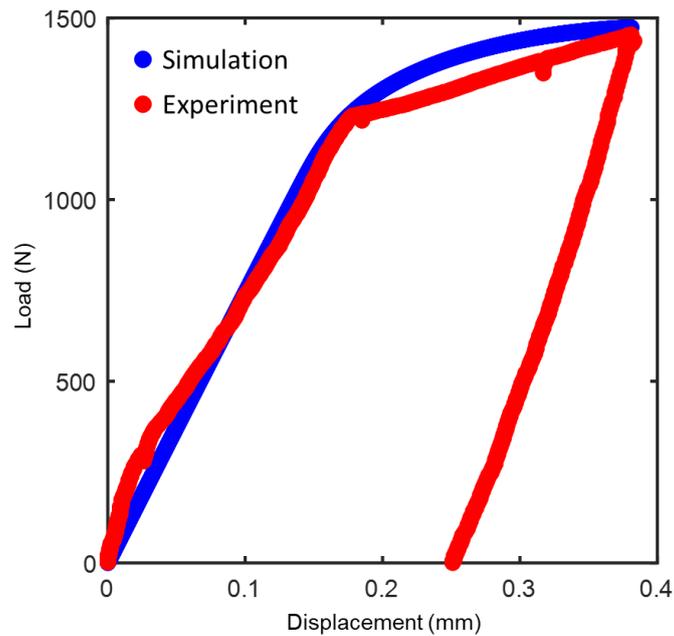

*Supplementary Figure 4: Simulated load-displacement show reasonable agreement with experiment.*